# Identification and characterization of various plastics using THz-spectroscopy

Tobias Kleinke[1], Finn-Frederik Stiewe[1], Tristan Winkel[1], Norman Geist[2], Ulrike Martens[1], Mihaela Delcea[2], Jakob Walowski[1*] and Markus Münzenberg[1]

[1]Institut für Physik, Universität Greifswald, Greifswald, Germany

[2]Institut für Biochemie, Universität Greifswald, Greifswald, Germany

*Author to whom correspondence should be addressed: jakob.walowski@uni-greifswald.de

**Abstract**. THz spectroscopy has reached imaging capability with a spatial resolution of a few micrometres. This property enables measurements and imaging of biological samples like cells. THz photons have very low photon energies in the millielectronvolt range. These energies interact mainly with vibrations in the molecules. Therefore, the investigated samples are neither ionized nor their DNA damaged, making this the perfect method for application on living cell tissue. In the future, material databases with absorption spectra will empower this method to distinguish even artificial particles like microplastics inside biological tissue. Current research aims at the impact of plastic nanoparticles on cell tissue considering various aspects. One of them is the investigation of the potential harm, the interaction of these particles with human body cells can cause, as these highly abundant particles are found anywhere in the environment. THz absorption spectroscopy offers the opportunity to investigate the polymers and identify them using their specific absorption lines as a fingerprint. In this publication, we study different polymers and analyse their particular fingerprints in the THz frequency spectrum from 0.1 up to 4 THz using a commercially available THz spectrometer. With this work we are contributing to the development, using THz waves for imaging, by identifying the fingerprints of the four most used plastics by humankind. At the end of this development, we envision a method to investigate the influence of nanometre-sized plastic particles on the human body and other biological organisms.

## 1. Introduction

Plastics are a commonly used material worldwide. The advent of industrialization and urbanization [1, 2] led to a consumer society in the course of the past century [3, 4]. This development created a strong demand for cost-effective and much lighter alternatives to common packaging materials like metal, glass or ceramics for the whole scale of retail goods [5, 6]. Because of their flexibility, the usage of plastics during this time increased progressively. The combination of low-cost and variable production makes them indispensable for industrial packaging [5, 7]. Due to their robustness, plastic packages offer a very good protection against external influences. The impermeability for gases, e.g. oxygen and carbon dioxide and the protection against moisture and light, makes them convenient for the food industry [7] as well as in other areas such as the construction or the automotive industry. However, this high demand leads to a lot of plastic waste around the globe. Only about 20% to 25% of the produced plastic is determined for long-term use and approximately 50% is disposable packaging [5, 8, 9]. Between 1950 and 2015, 8.3 billion tonnes of plastic were produced [5], 56% of that from the year 2000 onwards. Around 75% of that has now been disposed of as waste [10]. Less than 2% of the waste was recycled, around 6% incinerated and most of it, about 92%, was landfilled or disposed of anywhere in the nature [5, 9]. Around 4.8 to 12.7 million tonnes of the waste ends up in the oceans each year [11, 12]. Simulations estimate that by 2050 there will be more plastic waste in the oceans than fish in terms of weight [13]. This plastic decomposes into smaller and smaller parts down to micrometre sized particles, the so-called micro plastic, which is distributed throughout the environment [10, 12]. The ingestion of plastic and microplastics causes major problems for wildlife. It is estimated that by 2050 approximately 99% of all seabirds bodies will contain plastic [14]. Even the smallest microorganisms absorb plastic particles. The interaction between an organism and a plastic depends on the type of the ingested materials polymer composition [15]. A reliable detection for the respective plastic types, requires the development for identification and characterization methods. Terahertz radiation (THz) based methods, the so-called THz-spectroscopy [16] offer non-invasive investigation of biological cell tissue and aggregations therein. In the last few decades, the interest in the THz-frequency range increased in the fields of physics, chemistry, and material sciences. Due to its non-absorbent nature, e.g., in clothing, paper, wood or skin, terahertz radiation offers great potential for many applications. The fact that metal and liquids do not transmit this frequency makes it very interesting for airport security checks, medical imaging or spectroscopic applications [17]. The low photon energy enables a non-destructive and contactless examination and identification of materials and objects. In contrast to X-rays, cells or human tissues are not ionized or modified when exposed to THz-radiation. By using near-field optics, it is possible to achieve a very low spatial resolution of the radiation below the Abbe limit in the micrometre range [18]. Thus, it is possible to use THz radiation in various fields of medicine, for example for imaging in oncology or dermatology to detect cancer cells [19–21]. THz radiation can also be used for spectroscopy. The specific energy range induces rotational and vibrational transitions in molecules [17], which shows in the transmission spectra. Thus, it is possible to detect and identify substances such as drugs or explosives [22, 23] or to measure the humidity in air or objects [24]. Radiation can also be used to check the quality of medicines or food without unpacking [22, 25, 26]. The combination of the low photon energy and the possibility of a spatial resolution in the micrometre range [18] makes it interesting for the imaging and identification of foreign objects in cells.

In this study we investigate the behaviour of four commercially available plastics in the THz-frequency range. Polyethylene low density (PE-LD), Polyethylene high density (PE-HD), Polypropylene (PP) and Polystyrene (PS) are all used in common everyday lives. We find PE-LD for example in refuse bags, garbage bags or squeeze bottles. While PE-HD is used for shopping bags, buckets, milk bottles or shampoo, PP can be found in lunch boxes, packaging tape and garden furniture. PS is included in CD cases, protective packaging and food insulation [13]. In comparison to other experiments [16, 27–29] we use radiation with a higher wavelength between $600\,\mu m$ and $75\,\mu m$, which corresponds to a frequency range from $0.5\,THz$ to $4\,THz$. This frequency range is located in the gap between Fourier transform infrared spectrometry (FTIR), which use radiation between $15\,THz$ and up to about $120\,THz$ [16, 27, 28], and microwave spectroscopy with a frequency between $2.4\,GHz$ and $3.6\,GHz$ [29]. Due to the corresponding energy, radiation in the THz range signatures differ from FTIR measurements. THz-radiation induces lattice vibrations, so-called phonon vibrations, in crystal lattices [30, 31]. Hydrogen bonds and dipole-dipole bonds also interact with THz radiation [32]. These interactions with THz frequencies cause photon absorption and enable the identification of bonds, atoms, and materials.

## 2. Experimental setup

### 2.1 Terahertz spectroscopy

We use a standard THz spectrometer (TeraK 15, MenloSystems) to investigate the absorption of THz radiation in polymer samples. The system consists of a femtosecond fibre laser source emitting at $1.56\,\mu m$ central wavelength, which irradiates a photoconductive antenna (PCA), used as a THz-emitter. This PCA consists of 2 conductor tracks deposited on an LT-GaAs substrate. The conductor tracks are separated by a small gap consisting of the LT-GaAs layer. Applying a voltage to the antenna tracks does not lead to any current due to the high resistance in the gap area. However, applying a femtosecond laser pulse induces hot electrons in the LT-GaAs with lifetimes in the picosecond range and generates a current pulse in this area. The movement of charges acts as a Hertzian dipole emitting a THz pulse. A fibre-coupled optical light path with a delay line makes it possible to detect the THz-radiation with a second PCA, used as a THz-detector. No voltage is applied to the detector, however, when the THz pulse from the emitter arrives, its electric field determines the current generated by illuminating the gap. The delay line allows for a scan width of $> 850\,ps$, which leads to a spectral resolution up to $1.2\,GHz$. The optical light is delayed so that the THz pulse can be sequenced by the optical light pulse. The THz-spectrometer setup allows to focus the THz pulses on to the samples and thus measure the absorption reliably in the range from $0.5 - 4\,THz$. A Fourier transform of the THz pulses provides the spectral composition of the measured signal. During the measurements, the system is placed in a box constantly purged with nitrogen. This reduces the humidity to $< 1\%$ and the impact of water absorption from the normal laboratory environment in the measurements are minimized.

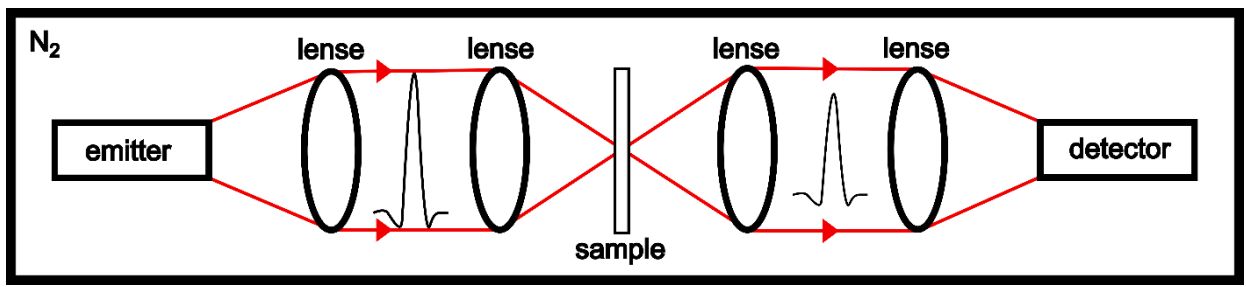

**Figure 1: Schematic diagram of the THz spectroscopy setup. The black box around the system indicates the airtight enclosure filled with pure nitrogen to eliminate humidity. The THz pulse generated by the emitter (optical path indicated by the red lines) is in a frequency range between 0.1 THz and 4 THz. The sample is placed in the beam focus behind the second lens. The transmitted radiation is measured by the detector and recorded using an amplifier.**

### 2.2 Samples

We study the characteristics of four different plastics that are common in everyday lives. The samples are small sheets with a defined thickness between $0.5 - 20\,mm$. We chose two sorts of Polyethylene one with high density and one with low density. They consist of repeating $C_2H_4$ units and differ based on their polymer chain branching. Whilst PE-HD has light branching and therefore a high density ($0.94\,\frac{g}{cm^3}$ to $0.97\,\frac{g}{cm^3}$), the chains in PE-LD are heavily branched and the density ($0.915\,\frac{g}{cm^3}$ to $0.935\,\frac{g}{cm^3}$) is therefore reduced [33]. The third sample is Polypropylene. It is produced by chain-growth polymerization of $C_3H_6$ units and consequently consists of repeating $C_3H_6$ units. The last sample, in this study is Polystyrene. It consists of repeating units of monomers of the aromatic hydrocarbon styrene ($C_8H_8$). All these plastics belong to the group of thermoplastics. These can be deformed by heating and become solid again when they cool down. The molecules of most thermoplastics such as polyethylene or polypropylene are bound by strong van der Waals forces and covalent bonds. This type of binding absorbs radiation in the low THz frequency range, so that an absorption characteristic can be expected in the transmission spectrum at certain frequencies.

2.3 FTIR spectra simulations

The FTIR spectra are obtained using the CP2K [34] simulation software and the TRAVIS [35] analyzation tool mainly following the procedure described by Thomas et al. [36]. Initial atactic polystyrene mono, tetra- and decamer chains were built using the Polymer Builder [37] of CHARMM-GUI [38] with terminal hydrogen capping.

## 3. Results and discussion

The investigated samples are characterized using commercial standard FTIR (PerkinElmer, Spectrum 65, FT-IR Spectrometer). Figure 2 shows the FTIR measurements of the investigated plastics in red together with the corresponding library spectra in black. The plots show the spectral transmission with high absorption at characteristic frequencies for the investigated materials. All measured samples have the same thickness of 1.5 cm to ensure equal conditions for relative absorption. Comparison of the experimental data with the library data, delivers the composition and purity of the examined samples. The measurements confirm over 90% material pureness with correlation factors of 0.984 (PE-LD), 0.976 (PE-HD), 0.964 (PP) and 0.933 (PS). The available library data does not differentiate between PE-HD and PE-LD, as the characteristic absorption frequencies are identical. However, the transmission amplitudes differ due to the polymer density variations between both materials. Table 1 lists the characteristic absorption frequency peaks and their excitation origins for the selected plastics. Usually, the absorption peaks mirror photon energies which match excitations at the bindings between carbon and hydrogen atoms, but other excitations are also possible.

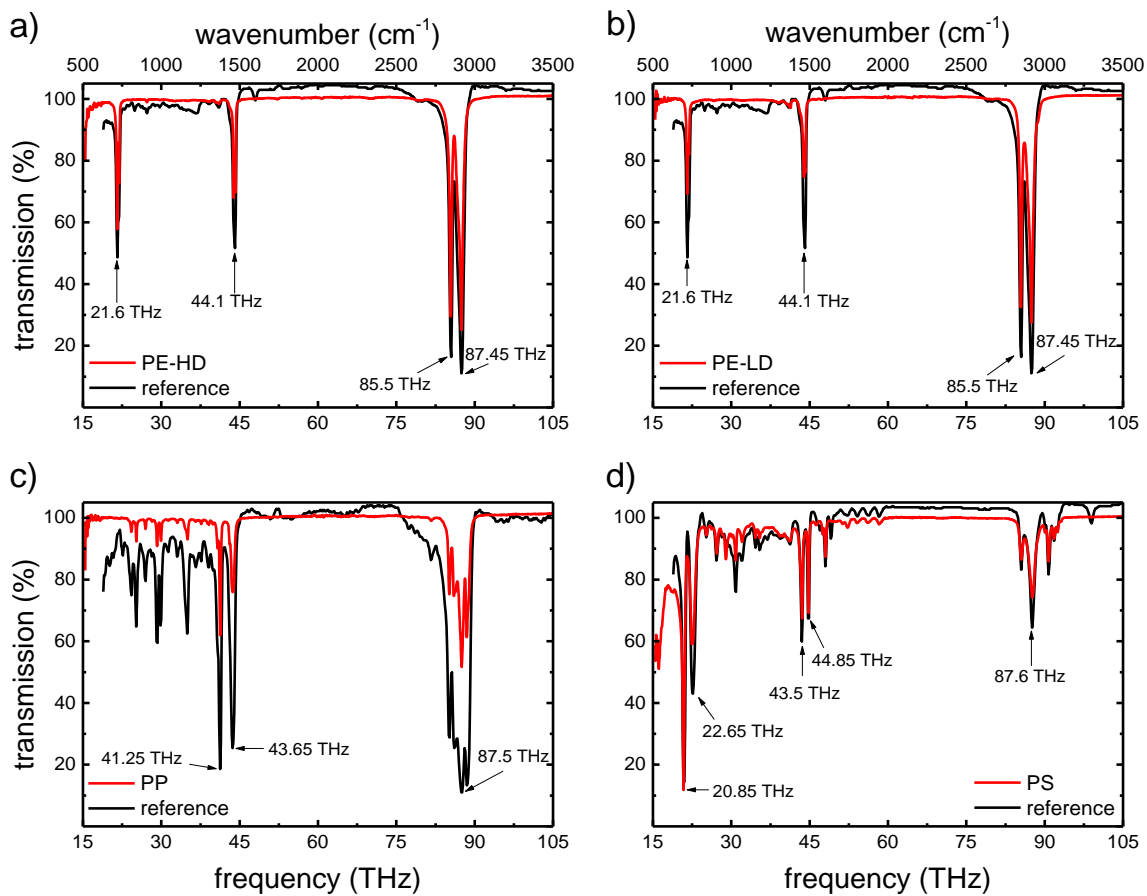

**Figure 2: FTIR frequency spectra for the investigated plastic samples. The red lines show the experimental data and the black lines the library data. All profiles within the study match the library data with correlation factors above 0.9.**

Table 1: Absorption peak frequencies and their origin. In general, the absorption frequencies match the excitation energies for the bonds between carbon and hydrogen or carbon and carbon atoms.

| Polyethylene [39] | | Polypropylene [40] | | Polystyrene [41] | |
|---|---|---|---|---|---|
| 21.6 THz | $CH_2$ rocking deformation | 41.25 THz | $CH_3$ symmetric bending vibration | 20.85 & 22.65 THz | C-H out-of-plane bending vibration |
| 44.1 THz | Bending deformation | 43.65 THz | $CH_3$ asymmetric bending vibration | | |
| 85.5 THz | $CH_2$ symmetric stretching | 85.5 THz to 88.65 THz | Symmetric and asymmetric C-H stretch | 43.5 & 44.85 THz | C=C stretching vibration |
| 87.45 THz | $CH_2$ asymmetric stretching | | | 87.6 THz | $CH_2$ asymmetric stretching |

For a better understanding of the absorption signals we add DFT based simulations for different chain lengths in increasing order to the measured FTIR spectra of polystyrene in figure 3. The computational expense increases disproportional to simulate the absorption band in the region between 0.3 and 5 THz, because the simulated system is not large enough and the timescale not suitable for this simulation method to resolve these slow modes sufficiently.

However, the simulations reveal, that even for short chains the characteristic absorption peaks appear in the infrared region and increase with the chain length, as is elucidated in the inset of figure 3. We expect that real samples contain a large range of chain lengths up to hundreds of monomers. Further, longer chain lengths exhibit additional vibrations that may arise from out-of-chain interactions when the polymers reach lengths which enable folding. This introduces modes that will also heavily appear in the bulk phase of the material and shows in the appearing of additional peaks and the increasing absorption background around those peaks. A comparison with experimental data shows, that for real polystyrene polymers consisting of long chains this background is even stronger, and several peaks shift to lower frequencies. Therefore, we also expect a peak broadening and smaller excitations in the THz range experimentally.

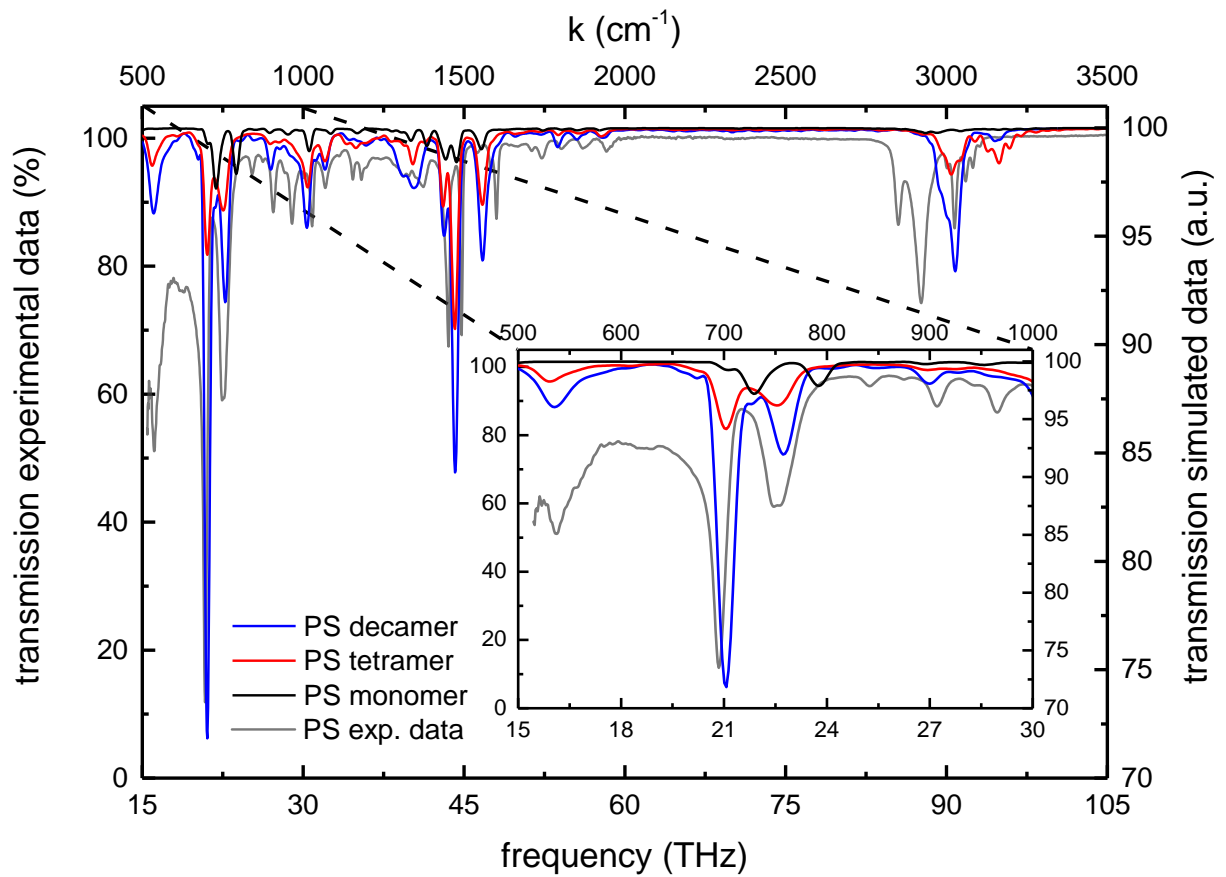

**Figure 3:** Simulated IR absorption spectra for polymer chains contained in bulk polystyrene compared to experimental data. Starting with a single molecule and increasing the chain length up to 10 molecules (decamer), the simulation data shows the absorption peak development. The inset elucidates, that longer chains allow for more excitations, increasing the absorption and forming more characteristic peaks, and thus, the correlation with the experimental data enhances for longer chains.

In the next step, THz absorption is investigated using the THz spectrometer. Figure 4 a) shows the reference measurement in nitrogen environment without a plastic sample in grey colour. By inserting the polymer samples into the beam path of the THz radiation, the pulses exhibit a time delay, and the signal amplitudes decrease slightly.

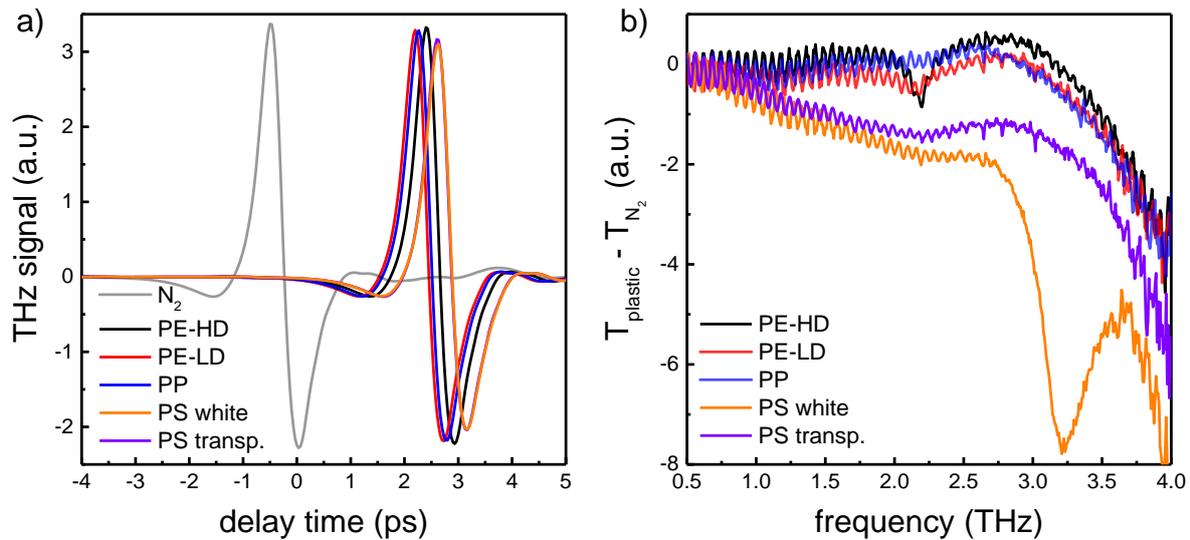

**Figure 4: a)** THz Electric field pules before (grey) and temporally delayed after passing different plastics samples with a thickness of $1.5$ mm. The plastics are mostly transparent to THz radiation, therefore, the transmitted amplitude through the samples does not change significantly compared to the initial pule. Further, the similar time delay by $\sim 3$ ps shows only a slight variation in the refractive indices. **b)** Fourier transformed spectra of the transmitted THz pulses $T_{\text{plastic}}$ in reference to the signal without a sample in the path $T_{N_2}$. Note the absorbance peak at $2.2$ THz for Polyethylene (black) and the increased overall absorption for the transparent Polystyrene. In contrast, the white PS sample shows a broad absorbance peak at $3.2$ THz stemming from the dye composition.

The delays of the THz pulses depend on the refractive index and the thickness of the inserted sample. Due to a close time delay and equal thickness of the investigated samples, their refractive indices have very similar values. The slight decrease in the amplitude of the transmitted THz radiation shows that all samples absorb and reflect only small parts of the THz radiation. The highest absorption in the THz frequencies can be seen for both PS samples, which decreases the amplitude of the THz signal by about 9%. The colour of the sample in this case has no influence on the delay of the THz pulse. This means that the refractive index of coloured and transparent PS samples is very similar. In general, these plastics are very transparent and suitable for THz optics like lenses.

Figure 4 b) shows the frequency spectra obtained by applying Fast Fourier Transformation (FFT) to the measured THz signals, plotted in figure 4 a). Characteristic absorption peaks are visualized, by subtracting the reference spectrum ($T_{N_2}$) from the frequency spectrum that remains after absorption by the plastic ($T_{\text{plastic}}$). Those spectra provide information, about the specific frequencies absorbed by the respective material. However, in contrast to the presented FTIR data, the 100% transmission signal for each frequency corresponds to the value 0 on the $T_{\text{plastic}} - T_{N_2}$ scale, while the negative values represent a decrease in transmission. The photon energies in the THz spectroscopy setup are about 10 times lower, than those in the FTIR spectrometer. However, it is possible to excite certain bonds or modes increasing the absorption of the corresponding frequency in the material. The graphs in figure 4 b), are plotted for a frequency range from $0.5 - 4$ THz, as this is the range specific for the emitter and detector material. The optimum frequency range for this spectrometer is $0.5 - 3$ THz. There, the detected amplitude is high and is decreasing strongly above $3$ THz. We can identify a clear absorption peak for PE-HD (black) at around f = 2.2 THz.

PE-LD (red) also absorbs this frequency, but the peak is not as clear due to its lower density and thus fewer molecules within the 1.5 mm plastic sheet to absorb photons at this frequency. The same effect is observed in the FTIR measurements, the absorption dips in figure 2 a) are stronger than those in figure 2 b). The photon energy in the infrared spectroscopy measurement of the absorbed THz radiation induces lattice vibrations in the PE samples [42–44] reducing the transmission amplitude at those frequencies, which are material specific for Polyethylene. In the PP sample, the molecule excitations in the infrared spectra appear already weak compared to the reference, shown in figure 2 c). Therefore, we also expect smaller excitations in the THz range and hence, the measured spectrum shows no characteristic absorption peak (figure 4 b).

The last two common plastics investigated in this study are two different types of PS, a transparent and a white coloured sample. The transparent sample shows a higher overall absorption than PE or PP but no specific absorption peak, which is consistent with other publications [43]. When dye is added to the sample the absorbance over the whole spectra increases and around 3 THz a broad absorption peak appears. Photon frequencies around 3 THz excite hydrogen bonds of OH groups [43, 45]. This leads to a higher photon absorption in this frequency range, especially at $f = 3.2\,\text{THz}$ a low transmission signal is detected. Figure 4 b) shows a broad $0.5\,\text{THz}$ peak with decreased transmission for the coloured PS. Differently coloured samples show the same absorption characteristics regardless of the plastic type, see **supplementary material**. This is a strong hint for specific absorption from the dye, because many of the commercially used colours contain OH-groups in their chemical structure.

The apparent absorption increase appearing for all measured materials at frequencies above $f = 3\,\text{THz}$ results from the detection threshold and reduced amplitudes for the respective frequencies. The signal absorbed by the plastics in this range results in a lower signal to noise ratio. The presented data already allows to differentiate between those three kinds of plastic materials, making a starting point for a materials database.

To verify the results presented in figure 4, we examine the thickness dependence of the transmission through the plastic for two polymer chains, PE-HD and PS. By implementing the PE-HD sample, the delay of the THz signal increases linearly for increased PE-HD thickness, shown in figure 5 a), as expected. The highest time delay can be seen for $20.01\,\text{mm}$ PE-HD, resulting in a time shift of about $35.5\,\text{ps}$. In general, it can be observed that the delay changes by about $1.77\,\text{ps}$ per 1 mm. This corresponds to refractive indices $n \approx 1.5 - 1.6$ for the considered frequency range in those materials. Figure 5 b) shows the transmission spectra for polyethylene at 5 different thicknesses. By increasing the sample thickness, the transmission of the THz radiation decreases at the frequency of $f = 1.2\,\text{THz}$. Also, the absorption peak at a frequency of about $f = 2.2\,\text{THz}$ gets more distinct for higher thickness because there are more molecules, which can absorb the radiation. This data confirms characteristic absorption peak for PE-HD. Implementing the PE-HD samples in the beam path and their absorption reduces the THz amplitude and the detectable signal already decreases below the threshold value of $-80\,\text{dB}$ for frequencies from $f = 2.8\,\text{THz}$ (for $20\,\text{mm}$ PE-HD) onwards. Smaller intensity values cannot be detected by this setup arrangement. This value provides information about the detection ability of the signal and the cut-off frequency, at which the signal is completely absorbed (for explanations see supplementary material).

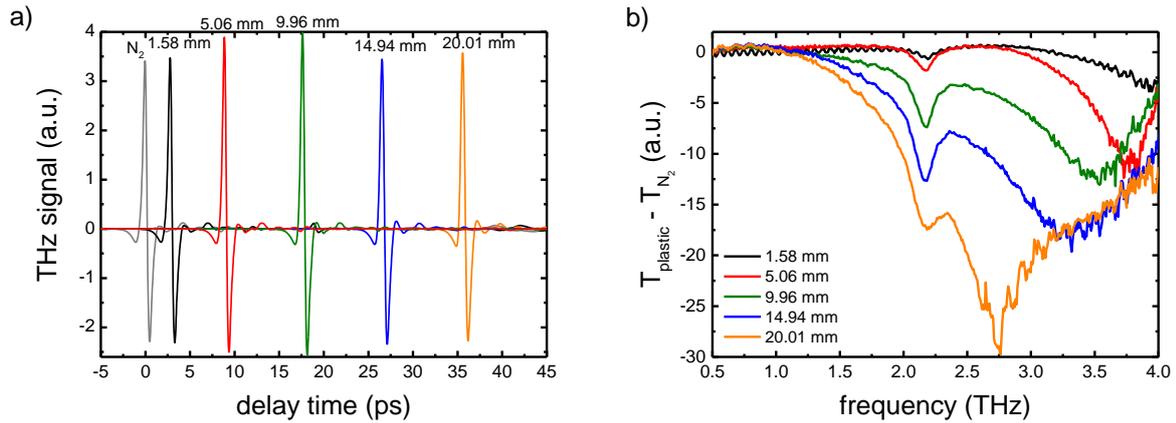

**Figure 5: THz transmission through PE-HD with increasing material thickness. a)** The temporal delay linearly increases with the PE-HD thickness about $1.77\,\text{ps/mm}$ according to a refractive index $n = 1.54$. **b)** Fourier-transformed transmission spectra of the data in a) after subtraction of the reference signal. The characteristic peak at $f = 2.2\,\text{THz}$ increases with the PE-HD thickness until the overall transmission decreases to around the detection limit for the $20.01\,\text{mm}$ sample.

Similarly, the same set of thickness dependent measurements for coloured polystyrene is depicted in figure 6. Figure 6 a) shows the recorded time domain data, an increasing delay with the sample thickness, which is larger than for PE-HD and indicates a slightly higher refractive index. The delay increases by about $\sim 2\,\text{ps}$ per $1\,\text{mm}$ PS.

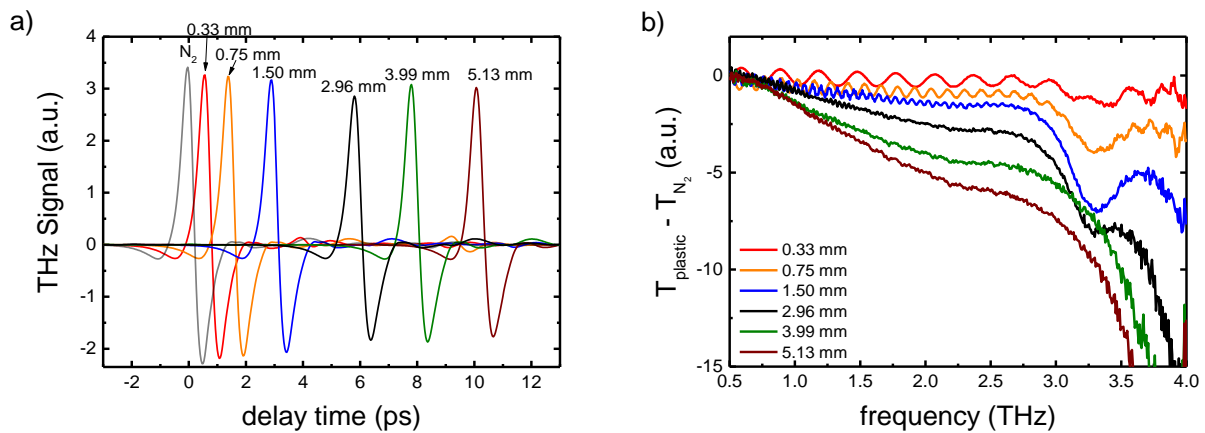

**Figure 6: THz transmission through coloured PS with increasing thicknesses. a)** The temporal delay of the THz pulses linearly increases with the PS thickness about $2\,\text{ps/mm}$ according to a refractive index $n = 1.59$. **b)** Fourier-transformed transmission spectra of the data in a) after subtraction of the reference signal. The overall absorption increases with the sample thickness, especially for the higher THz frequencies. At $3.2\,\text{THz}$ the broad characteristic absorption peak appears, which stems from hydrogen bonds in OH-groups in the dye additive.

Figure 6 b) shows the FFT of the THz signal with PS in the beam path from which the FFT without PS is subtracted. Increasing the thickness of the PS sample causes an increase in the absorption of the THz radiation. For the thicknesses $0.33\,\text{mm}$, $0.75\,\text{mm}$, $1.7\,\text{mm}$ and $2.96\,\text{mm}$ an additional absorption at a frequency of $f = 3.2\,\text{THz}$ appears and increases for higher thicknesses. Above $2\,\text{mm}$, the absorption peak is not recognizable anymore, because the overall intensity of the frequencies in the transmitted signal is below the threshold of $-80\,\text{dB}$, which is the smallest value the system can detect. In general, we expect a stronger absorption with increasing material thickness, since more molecules, bonds or lattice vibrations are excited.

## Conclusions

In summary, we have demonstrated a method to identify various plastics using THz radiation. We find a characteristic absorption band for PE-HD ($f = 2.2\,\text{THz}$), and an increased overall absorption for PS. Besides this, we identify an absorption band for PS and other plastics ($f = 3.2\,\text{THz}$) if colour additives based on OH groups are added. The absorption peaks for PP in this frequency range cannot be distinguished clearly by this method. The identified characteristics can be further confirmed by thickness dependent measurements of these plastics. The absorption peaks remain at the same frequency, but broaden, as more bulk interaction is available for the photons. With increasing thickness, also the overall absorption increases improving the visibility of the characteristic absorption peaks, but as the overall absorption diminishes the THz signal, the peaks are not detectable anymore after a certain threshold. In addition, we can determine that all the examined plastics are largely transparent to radiation in the THz frequency range. Our work serves as a basis to represent and identify these substances in biological cells or tissue. For this study we measured samples in the millimetre range to obtain characteristics for the absorption. The material identification is supported by measurements and simulations in the FTIR frequency range. The measurements confirm the materials purity, and the simulations show, how the absorption peaks develop with the polymer chain lengths. Further studies of plastic sediments in biological cell tissue to detect particles in the sub micrometre size are necessary. The development in the field of THz spectroscopy already allows to distinguish micrometre sized particles using near field imaging [18, 46–48]. Other articles introduce methods to amplify the THz signal and at the same time achieve sub micrometre resolution [49]. In this work we make the first move to create a database for characteristic absorption spectra in the THz range of polymers common in everyday lives. Combining the obtained characteristics with the near field imaging approach and other available methods will allow to enhance the detection limit and image particles in the sub micrometre range.

# Conflict of Interests

The authors have no conflicts of interest to disclose.

# Acknowledgements

The authors gratefully acknowledge the financial support from the BMBF, MetaZIK PlasMark-T (No. FKZ:03Z22C511).

# Identification and characterization of various plastics using THz-spectroscopy

Tobias Kleinke[1], Finn-Frederik Stiewe[1], Norman Geist[2], Ulrike Martens[1], Mihaela Delcea[2], Jakob Walowski[1*] and Markus Münzenberg[1]

[1]Institut für Physik, Universität Greifswald, Greifswald, Germany
[2]Institut für Biochemie, Universität Greifswald, Greifswald, Germany
*Author to whom correspondence should be addressed: jakob.walowski@uni-greifswald.de

## Absorption characteristics for color additives

The transmitted THz signals and the corresponding frequency spectra of PP, PS and PVC with and without color additives in a frequency range between $0.5\,\text{THz}$ and $4\,\text{THz}$ are depicted in figure S1. The THz-signals are measured with a commercial standard THz spectrometer (TeraK 15, MenloSystems). Implementing the plastics in the beam path leads to a time delay in the $\text{ps}$ range, see figure S1 a). The detected signals for the plastics without color additives (solid lines) show a larger temporal delay as the colored plastics (dashed lines). This comes from the smaller sample thicknesses of the samples containing color additives and does not refer to different refractive indices. The investigated sample thicknesses are listed in Table 1.

*Table S1: Thickness of the investigated plastics samples for color additives dependency.*

| Material | Thickness in mm |
|---|---|
| PP transparent | 1.61 |
| PP black | 0.82 |
| PS transparent | 1.54 |
| PS white | 1.50 |
| PVC transparent | 0.95 |
| PVC red | 0.318 |
| PVC blue | 0.295 |

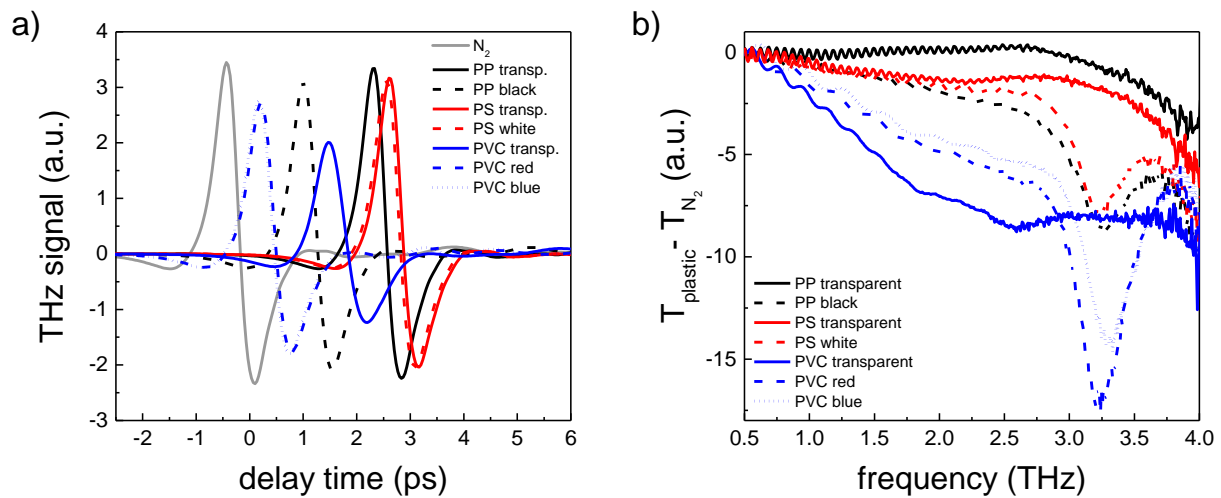

**Figure S1: a) THz field after absorption by different plastics placed in the beam path. Three of the used samples are transparent and without color additives (solid lines). The other materials contain additional dye. The transmission amplitude of PVC is lower than for PP or PS, indicating a stronger absorption or reflection by this material. (b) Fourier transformed spectra of the transmitted THz pulses $T_{plastic}$ in reference to the signal without a sample in the beam path $T_{N_2}$. The transparent plastics show no specific absorption pattern. However, all materials with color additives show a broad absorption peak at $3.2\,THz$.**

Figure S1 b) shows the extracted characteristic peaks in the Fourier transformed signal. For plastics without color additives, shown in solid lines (PP in black, PS in red and PVC in blue), there are no absorption peaks. For all colored samples a high absorption in the range from around 3.0 THz to 3.5 THz is detected. This behavior comes from the additional used color in those plastics, which in general contains OH-groups. These atoms are bonded by hydrogen bonds, which are excited by radiation in the absorbed frequency range.

# FTIR spectra simulation procedure

Each polymer chain is subjected to equilibration molecular dynamics (MD) simulation in vacuum with NAMD 2.13 [1] and the CHARMM36 [2] force field, without periodic boundary conditions. This is useful to relax especially the longer chains and thereby introduce self-interaction via folding of the polymer strand. Short-range interaction cut-offs are set to 1 nm with a switching function of 0.1 nm. The timestep is set to 1 fs and all bonds to hydrogen atoms are constrained. After 1000 steps of energy minimization, NVT conditions followed for 1 ns. Temperatures are controlled to 300 K with a Langevin thermostat at a 1 $ps^{-1}$ damping coefficient.

The resulting end-state is in the next step subjected to ab-initio (AI) equilibration and production MD simulation with CP2k and the QuickStep [3] approach with orbital transformation (OT) [4]. Density functional theory (DFT) is applied as the electronic structure method. The BLYP exchange-correlation functional is applied, along with the optimized double-zeta valence and single polarization basis set for molecules (MOLOPT-DZVP-SR-GTH) and Goedecker-Teter-Hutter pseudopotentials [5–7]. The plane wave cut-off is set to 280 Ry, with four grid levels and default reference grid cut-off. Temperatures are adjusted to 400K by a Nosé-Hoover chain thermostat at 100 fs time constant and the timestep for MD is set to 0.5 fs. The convergence threshold for the self-consistent field iterations is set to $1e^{-5}$ with a maximum of 100 iterations. The polymers chains are placed in periodic cube-shaped boxes of 8, 16 and 25 Å for the mono-, tetra- and decamer, respectively. The systems are subjected to single-point-optimization and subsequent geometry optimization for five steps. Heating and short equilibration follows for 1 ps under NVT conditions. The production trajectory is calculated under the same conditions for 100 ps.

Because volumetric electron densities are too large to store, the resulting AIMD production trajectory is post-processed with CP2k in chunks of 5000 frames and a stride of eight frames (4fs) to compute the volumetric electron densities stored in the CUBE format. These chunks are then processed with TRAVIS to compute and store the electro-magnetic properties (EMP) of the system with the Voronoi-tessellation approach [8]. Once this process is completed, TRAVIS is applied again to compute the final FTIR spectrum from resulting concatenated electro-magnetic moments of the system.